# Enhanced infrared vision by nonlinear up-conversion in nonlocal metasurfaces

**Laura Valencia Molina[1,2]\* | Rocio Camacho Morales[1]\* | Jihua Zhang[1†] | Roland Schiek[1] | Isabelle Staude[2] | Andrey A. Sukhorukov[1] | Dragomir N. Neshev[1]**

[1]ARC Centre of Excellence for Transformative Meta-Optical Systems (TMOS), Department of Electronic Materials Engineering, Research School of Physics, The Australian National University, Canberra, ACT 2600, Australia

[2]Friedrich Schiller University Jena, Institute of Solid State Physics, Max-Wien-Platz 1, 07743 Jena, Germany

**Correspondence**
Dragomir N. Neshev
Email: Dragomir.Neshev@anu.edu.au

**Present address**
[†]Songshan Lake Materials Laboratory, Dongguan, Guangdong 523808, P.R. China

**Funding information**
Australian Research Council, CoE program (CE200100010), the German Research Foundation (DFG, Deutsche Forschungsgemeinschaft) through the International Research Training Group Meta-ACTIVE (IRTG 2675, project number 437527638),

The ability to detect and image short-wave infrared light has important applications in surveillance, autonomous navigation, and biological imaging. However, the current infrared imaging technologies often pose challenges due to large footprint, large thermal noise and inability to augment infrared and visible imaging. Here, we demonstrate infrared imaging by nonlinear up-conversion to the visible in an ultra-compact, high-quality-factor lithium niobate resonant metasurface. Images with high conversion efficiency and resolution quality are obtained despite the strong nonlocality of the metasurface. We further show the possibility of edge-detection image processing augmented with direct up-conversion imaging for advanced night vision applications.

**KEYWORDS**

nonlinear metasurface, up-conversion imaging, lithium niobate

\*Equally contributing authors.

# 1 | INTRODUCTION

Advanced infrared (IR) vision has many important applications in areas such as surveillance, security, and medical imaging. The IR spectral range around 1550 nm is particularly useful for optical communications, LIDAR, and nightglow vision [1, 2]. However, traditional IR imaging devices that rely on narrow bandgap semiconductors, like InGaAs, are limited by low-temperature operation and high noise levels [1]. Therefore, it is beneficial to convert IR light into visible so that it can be detected with little noise by conventional silicon-based imaging sensors and potentially augmented the IR with visible imaging.

The technique of nonlinear up-conversion IR imaging [3] has shown significant promise in enabling such conversion. This technique is based on the parametric nonlinear process known as sum-frequency generation (SFG), which increases the energy of the incident signal photons by mixing with a plane pump beam. The technique is highly advantageous, as it enables a coherent state conversion (preserves the coherence of the incident wavefront). However, the large footprint of the existing apparatus due to the use of bulky crystals has limited its practical applicability.

To overcome these limitations, resonant dielectric metasurfaces have been recently explored as nonlinear up-converters of ultra-thin footprint. Metasurfaces are optical surfaces engineered to exhibit optical properties that are not found in conventional materials. They are composed of nanoscale building blocks, also known as meta-atoms, that are designed to manipulate the wavefront of light with sub-wavelength resolution. Local metasurfaces are intended to control the local scattering of the unit cells, leading to spatial control over the phase, amplitude and polarization. Local metasurfaces were explored to overcome the limitations of conventional imaging systems, resulting in achromatic meta-lenses with enhanced phase-compensation and efficiency [4, 5]. In nonlinear up-conversion IR imaging, resonant dielectric metasurfaces have been recently explored to increase the SFG conversion efficiency [6, 7, 8] due to the strong field enhancement at the metasurface resonant frequencies. The small thickness, lack of longitudinal phase matching, and efficiencies approaching those of bulk crystals have driven interest in this direction. Mie-resonant GaAs metasurfaces have been tested for IR imaging [7]. However, the limited quality (Q-) factors of the localized resonances and the high absorption of the GaAs material at visible wavelengths have limited the conversion efficiency and the overall performance.

A naive approach to increase the efficiency of the nonlinear conversion process is to employ high-Q-factor metasurfaces [9, 10]. However, the narrow linewidth of the high-Q resonances reduces the operation bandwidth [11]. Furthermore, the often strong spatial nonlocality (angular dispersion) of the resonances would limit the resolution quality of the images. It is, therefore, a critical challenge to derive imaging schemes that enable both high conversion efficiency and image quality. Here, for the first time to our knowledge, we demonstrate IR imaging using a nonlinear up-conversion process in a nonlocal high-Q lithium niobate (LiNbO$_3$) metasurface. By appropriate Fourier imaging, we overcome the limitations of spatial nonlocality and demonstrate record-high conversion efficiency and image quality. Furthermore, we propose a novel functionality of the metasurface where the up-conversion process can simultaneously perform image processing, such as edge detection, together with direct IR imaging in the different diffraction channels of the metasurface.

# 2 | UP-CONVERSION IMAGING PRINCIPLE

The concept of infrared vision by nonlinear up-conversion is depicted in Figure 1(a). Under this scheme, the infrared light transmitted (or reflected) from an object (depicted as a kangaroo in the figure) is imaged onto a thin layer that converts all the features of the object to a new shorter-wavelength visible light. In the ideal case, where there is no loss of information on the converted wavefront, this visible up-converted image can be detected with a conventional camera. The up-conversion screen is the key component of such an infrared imaging scheme, as it needs to provide additional energy to the incoming IR photons while preserving the wavefront information. Several different up-conversion schemes have been explored, including screens based on up-converting rare-earth nanoparticles [12], organic semiconductors heterojunctions [13] and nonlinear sum-frequency generation [3]. Out of these different mechanisms, the parametric process of sum-frequency generation (SFG) is the only mechanism that fully preserves the properties of the IR wavefront while frequency up-shifting the energy of the input wave. As such, despite the lower quantum efficiency of the SFG process, it finds practical important applications. First, the up-conversion screen can be inserted at any position in the imaging system, and for ultra-thin screens (such as metasurfaces), it enables minimal changes in the imaging system. Second, as the output wavefront is preserved, it can be further analyzed for phase and polarization detection [14] or other image-processing applications [15, 16]. Finally, the SFG process is ultra-fast and enables femtosecond time resolution, thereby removing any time lag in the imaging process.

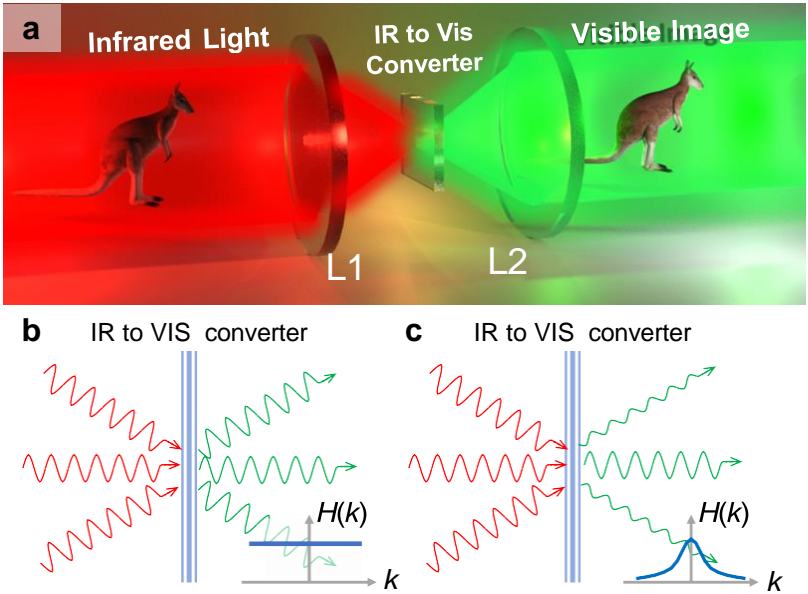

**FIGURE 1**  Infrared (IR) to visible (VIS) up-conversion for vision applications. (a) Schematic of the nonlinear up-converter for infrared imaging, where infrared light illuminating an object and passing through a lens (L1) is coherently up-converted to visible light and captured by another lens (L2) to be finally observed on a conventional silicon-based camera. (b) The ideal up-converter shall convert all rays, incident at different angles, with the same efficiency, i.e. $H(k)$ = constant. Here, $H(k)$ is defined as the up-conversion transfer function. (c) In practice, due to the spatial dispersion (or nonlocal properties) of the up-converter, the components propagating at normal incidence are converted with higher efficiency than the ones propagating at larger incident angles, i.e. $H(k_{low}) > H(k_{high})$.

These advantages of the SFG process as an up-converter can be understood by writing the SFG electric field, $E_{\omega_3}(\zeta, \eta)$, on an ultra-thin up-converter, including a nonlinear metasurface, as

$$E_{\omega_3}(\zeta, \eta) \sim \varepsilon_0 \chi^{(2)} E_{\omega_1}(\zeta, \eta) E_{\omega_2}(\zeta, \eta), \tag{1}$$

where $(\zeta, \eta)$ are the transverse spatial coordinates on the up-converter, $E_{\omega_{1,2}}(\zeta, \eta)$ are the electric fields of the required pump (1) and IR signal (2) beams on the metasurface, $\varepsilon_0$ is the dielectric permittivity of vacuum, and $\chi^{(2)}$ is the quadratic nonlinear susceptibility tensor of the metasurface material. For simplicity, here we assume that $\chi^{(2)}$ is a scalar value. This assumption is valid in the case of $x$-cut LiNbO$_3$ material, where the $\chi_{zz}^{(2)}$ is the dominant tensor component. The fields, in this case, are linearly co-polarised along the optical axes of the crystal. When the field on the up-converting screen is obtained by a single lens, L1 in Figure 1(a), the field is the Fourier transform of the input image, $E_\omega = \mathsf{F}\left(\bar{E_\omega}\right)$, where $\bar{E_\omega}$ is the image at the front focal plane of the lens. The SFG field on the metasurface then can be written as

$$E_{\omega_3} \sim \varepsilon_0 \chi^{(2)} \, \mathsf{F}\left(\bar{E}_{\omega_1} * \bar{E}_{\omega_2}\right),\tag{2}$$

where the symbol $*$ denotes convolution, while $\mathsf{F}$ denotes a Fourier transform. This important result shows that the SFG process is a convolution of the two input fields, $E_{\omega_1}$, $E_{\omega_2}$ (taken at the front focal plane of the lens). Such an up-conversion scheme can, therefore, be used to perform convolution operations and can be used for all-optical image processing [16]. In most vision applications, however, direct image conversion is required. In this case, the pump can simply be a plane wave or a broad Gaussian beam. Finally, another lens in the imaging arm of the system, L2 in Figure 1(a), can perform an inverse Fourier transform to produce a full-wave image of the object, but at an up-converted wavelength:

$$E_{\omega_{out}} = H \, E_{\omega_{in}},\tag{3}$$

where $E_{\omega_{out}} = \bar{E}_{\omega_3}$ is the output SFG field, $E_{\omega_{in}} = \bar{E}_{\omega_2}$ is the signal field, containing the information about the IR image, and $H$ is the transfer function that depends on the nonlinear susceptibility and the pump field $\bar{E}_{\omega_1}$.

In the general case of nonlinear frequency up-conversion, other nonlinear processes can also take role, such that the transfer function is nonlinear to the signal field, $E_{\omega_{out}} = H\, E_{\omega_{in}} + H^{(2)}E_{\omega_{in}}^2 + H^{(3)}E_{\omega_{in}}^3 + \cdots$, where $H^{(2)}$ and $H^{(3)}$ are the higher order transfer functions that can have a tensorial form (similar to the nonlinear susceptibility). Examples include imaging by second harmonic generation (SHG) and third harmonic generation (THG) [17, 18]. However, the transfer function is linear only for the SFG process, which is highly beneficial for vision applications. The SFG process further allows the use of weak signal beams, as the conversion can be boosted by the employment of a strong pump beam. We note that in the above imaging applications, the image information is encoded in the input beam, in contrast to works where the image is encoded in the transfer function, $H$ [19, 20, 21]. This latter case, however, is impractical for vision applications. Therefore, the development of an up-converter that is ultra-thin, highly efficient and exhibits a linear transfer function is a major challenge in the field of IR up-conversion.

Quadratic nonlinear metasurfaces have offered a promising solution, as they can significantly enhance nonlinearity due to their ultra-thin form factor and underlying optical resonances that promote near-field enhancement. Broadly speaking, optical metasurfaces exhibit two types of resonances [22]: localized Mie-type resonances and collective lattice resonances [23, 24]. While Mie-resonant metasurfaces have been explored for up-conversion imaging [7], their low Q-factor resonances offer only moderate up-conversion enhancement. Enabling up-conversion imaging with high-Q resonances promises significant advantages due to the strong nonlinearity enhancement [25, 26], including continuous wave operation [9]. However, the strong spatial dispersion of high-Q metasurfaces has inhibited such IR vision applications to date. Such dispersion is associated with strong nonlocality of the metasurface [27, 28], namely that the linear transmittance strongly depends on the transverse momentum of excitation,

$$E_{out}(k_\zeta, k_\eta) = t(k_\zeta, k_\eta)E_{in}(k_\zeta, k_\eta),$$ (4)

where $k_\zeta, k_\eta$ are the transverse wavevectors on the metasurface and $t$ is the complex transmission coefficient. Inevitably, this linear nonlocality of the high-Q metasurface is translated into their nonlinear properties [29], making the nonlinear frequency conversion strongly dependent on the transverse wavevector:

$$E_{\omega_{out}}(k_\zeta, k_\eta) = H(k_\zeta, k_\eta) E_{\omega_{in}}(k_\zeta, k_\eta),$$ (5)

where $k_\zeta, k_\eta$ are the transverse wavevectors of the input IR wave on the metasurface. Note that the input and output fields are at the different wavelengths, $\omega_{in} = \omega_2$ and $\omega_{out} = \omega_3$. The up-conversion transfer function $H(k_\zeta, k_\eta)$ is, therefore, dependent on the angular dispersion of the metasurface at the signal wavelength and the beam shape of the pump beam. In the ideal case, the up-conversion transfer function, $H(k_\zeta, k_\eta)$, should be independent of the transverse wavevector, see Figure 1(b). This is often the case for metasurfaces exhibiting localized resonances, such as in Mie-resonant metasurfaces. However, in the case of high-Q nonlocal metasurfaces, the frequency up-conversion transfer function, $H(k_\zeta, k_\eta)$, could quickly reduce in value with increasing $k$, see Figure 1(c). More intricate up-conversion functions could also be considered, which would lead to incomplete image transfer. In Figure S1 of Supporting Information, we show the example of a meta-grating metasurface [30]. In this case, depending on the operating wavelength, the transfer function could exhibit a single-band or double-band type shape, resulting in different image formations from the original object.

Therefore, the main challenge for up-conversion IR imaging screens is to convert all spatial frequencies without loss of information from the original image. A possible solution to this challenge is to use an optical imaging scheme, where the Fourier plane of the input image is projected onto the up-converting screen rather than the image plane. In this case, the spatial frequencies of the input image are distributed as spatial positions on the metasurface and uniformly converted from IR to visible. The drawback is that the different spatial positions of the image are now converted to different wavevectors on the metasurface, resulting in asymmetric image conversion with respect to the two transverse directions. However, this asymmetry is somewhat circumvented by the broader transfer function, as seen in Figure S1(b). Here, we combine such an imaging scheme with a high-Q nonlocal metasurface to demonstrate record-high image up-conversion from IR to visible with high spatial resolution.

## 3 | RESULTS

In our work, we utilize a $LiNbO_3$ metasurface, schematically shown in Figure 2(a). The metasurface consists of a periodic $SiO_2$ grating on top of a thin $LiNbO_3$ film of subwavelength thickness. The metasurface has an etchless design, where the thin $LiNbO_3$ film is not etched. This etchless design assures high optical fields inside the nonlinear materials and improves the nonlinear conversion efficiency [30, 31, 32]. The choice of $LiNbO_3$ metasurface is further enhanced by the high transparency of the material at visible and IR wavelengths, as well as the high nonlinear coefficient [33]. These unique properties have led to an increased interest in the use of $LiNbO_3$ for meta-optics applications [34], including second- [35, 36, 37, 38, 39, 40] and higher-harmonic generation [38]. Importantly, a number of studies have explored high-Q metasurfaces for nonlinear enhancement [41, 37, 39]. However, the process of SFG has only been reported theoretically [31], and no studies for its use in up-conversion imaging exist to date.

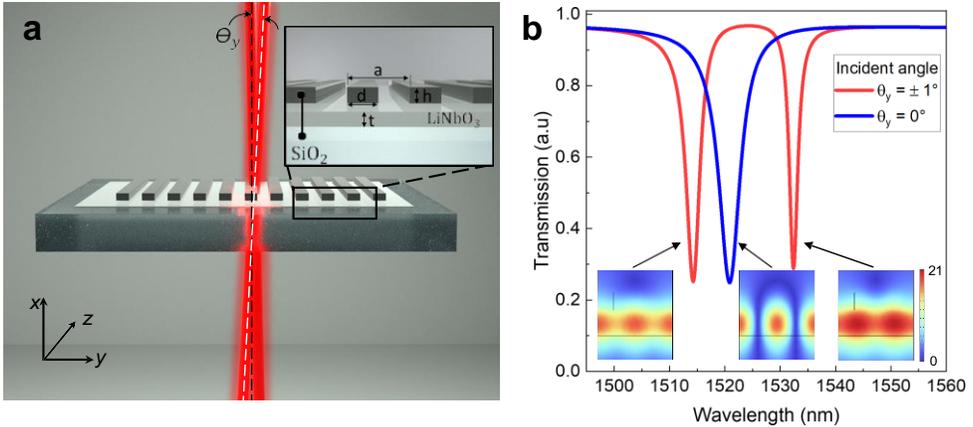

**FIGURE 2** Nonlocal metasurface geometry. (a) Geometry of the LiNbO₃ nonlocal metasurface, composed of a $t$ = 307 nm thin LiNbO₃ layer with a 1D SiO₂ grating on top. The geometric parameters of the grating are $a$ = 879 nm, $d$ = 500 nm, and $h$ = 200 nm. The designed metasurface shows a strong disperion when the illumination direction is tuned from normal incidence (black dotted line) to an angles $\theta_y$ along the grooves of the grating (white dotted line). (b) Simulated transmission spectra of the metasurface for normal ($\theta_y$ = 0° - blue line) and for $\theta_y$ = ±1° incidence (red line), showing the splitting of the main resonance due to its highly nonlocal properties. The bottom insets show the intensity distribution of the metasurface modes ($|E_z/E_0|^2$) at different resonant wavelengths.

## 3.1 | Linear nonlocal properties

The key geometric parameters of our metasurface are the LiNbO₃ layer thickness, $t$, the one-dimensional periodicity, $a$, and finally the width, $d$, and thickness, $h$, of the SiO₂ bars, as shown in the inset of Figure 2(a). The LiNbO₃ crystal is $x$-cut with $z$-axis aligned to the SiO₂ grooves. The operation of the metasurface is governed by the excitation of a guided mode inside the LiNbO₃ layer [42]. The coupling to this mode is controlled by the SiO₂ grating, resulting in a high-Q resonance. In the limiting case of $h \rightarrow 0$, the quality factor of the resonance theoretically approaches infinity. In practice, the fabrication imperfections give rise to more moderate Q-factors, which in our fabrication can vary from tens to a thousand. Figure 2(b-blue curve) shows the calculated resonance dip for geometric parameters $t$ = 307 nm, $a$ = 879 nm, $d$ = 500 nm, and $h$ = 200 nm at normal incidence. The calculated Q-factor of the resonance is $Q \approx 200$.

The physical mechanism of resonant excitation from guided modes spreading out in-plane underpins an intrinsically nonlocal response of the metasurface that is associated with strong angular dispersion [27, 43, 44]. As a result, even at a small angle of incidence, the resonance splits into two branches: one dark-mode branch at longer wavelengths and a bright-mode branch at shorter wavelengths. The dark mode is weakly coupled to free space and, as such, has a higher Q-factor. In Figure 2(b-red curve), we show the transmission spectrum of our metasurface for an incident angle of ±1°, considering a rotation axis along the grooves of the grating (see Figure 2(a), red beams). We see that even at this very small angle, there is a resonance split of about 20 nm. Such splitting affects the resonant enhancement of beams that exhibit non-zero transverse momentum and challenge imaging-type applications. In the inset of Figure 2(b), we also plot the spatial beam profiles of the different modes, where most of the energy is localized within the LiNbO₃ film, which is advantageous for the nonlinear conversion.

The metasurface is fabricated by thin film deposition of a SiO₂ layer, followed by electron beam lithography and a dry-etching process [30], see details in Sec. S1 of the Supporting Information. Several metasurfaces with different quality factors have been fabricated. The size of the fabricated samples is 400 $\mu$m. A scanning electron microscope (SEM) image of a typical metasurface is shown in Figure 3(a), where the inset shows a close-up of the SiO$_2$ ridges, demonstrating good uniformity with minimal surface roughness. The linear transmission spectrum of the metasurface is measured using a homemade transmission setup utilizing a white light source linearly polarized along the grooves of the gratings and Ocean Optics spectrometers (one for the visible and one for the IR spectral range). To measure the transmission spectra of the high-Q metasurfaces, we also used a tunable laser at the telecommunication range as the light source.

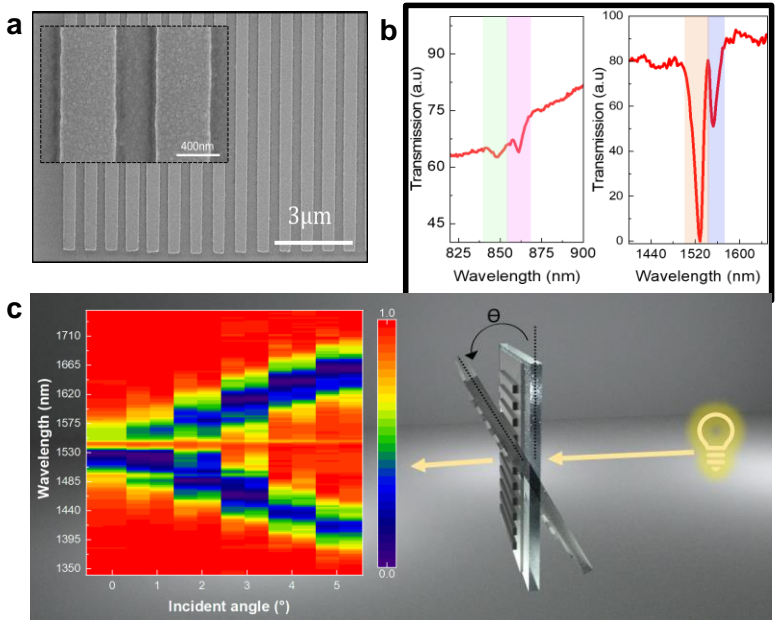

**FIGURE 3** Measured linear response of the fabricated nonlocal metasurface. (a) SEM image of the fabricated metasurface on top of an $x$-cut, 307-nm-thick LiNbO$_3$ film. The inset shows the 1D SiO$_2$ grating and the good quality of the side walls. (b) Linear transmission of the metasurface at normal incidence. The metasurface shows a resonant behavior around the pump wavelength at 847 nm and 860 nm. In addition, resonant behavior is also observed at the signal wavelength, at 1528 nm and 1552 nm. (c) Linear transmission as a function of incident angle $\theta$ and illuminating wavelength. There is a nearly degenerate resonance at normal incidence, followed by mode splitting at larger angles. The right panel shows the schematic of the measurement arrangement, where the metasurface is illuminated at different angles by a broadband tungsten-halogen lamp.

The transmission spectrum of a typical metasurface (corresponding to Sample 1 in Table S1) measured at normal incidence is shown in Figure 3(b). Two resonances are observed around 1530 nm with $Q \approx 40$. The large resonance dip at 1528 nm is the resonance expected at normal incidence, corresponding to the blue curve in Figure 2(b). The small resonance dip at 1552 nm corresponds to the dark mode, which becomes visible even at normal incidence due to fabrication imperfections. The resonances appear broader than in simulations, likely due to a number of factors, including fabrication imperfections, a slight etch on the LiNbO$_3$ film, and the resolution of our IR spectrometer (about 4 nm). In the near-IR spectral range ($800 - 900$ nm), see Figure 3(b, left panel), we also observe two other high-Q resonances. They, however, appear very shallow, likely due to the coarse resolution of our spectrometer (about

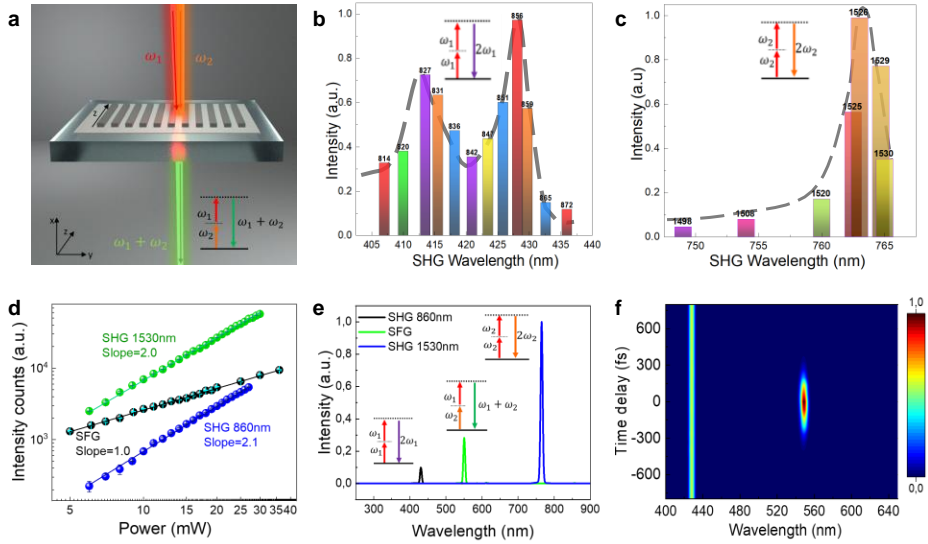

**FIGURE 4** Nonlinear characterization of the nonlocal metasurface. (a) Schematic of the metasurface infrared up-converter designed for sum-frequency mixing, as indicated by the energy diagram in the inset. (b) Measured nonlinear response of metasurfaces for pump beam. The wavelength of the pump is tuned from 820 to 875 nm, in steps of 5 nm. A maximum on the SHG is observed at 860 nm. (c) Measured nonlinear response of fabricated metasurfaces for signal beam. The pump wavelength is tuned from 1500 to 1540 nm, showing a maximum on the SHG at 1530 nm. (d) SHG intensity as a function of average power in the pump (blue) and signal beam (green). In addition, SFG intensity is shown as a function of average power in the signal beam (black). The experimental data, shown in a log-log plot, indicate a quadratic dependence of the second harmonic processes with the power of the pump and signal beams and a linear dependence of the SFG process with the power of the signal beam. (e) Measured intensity of the nonlinear emissions generated by metasurface (colored lines), corresponding to the SHG and the SFG processes. The corresponding energy level diagrams of the nonlinear processes are shown in every case. (f) 2D nonlinear spectrum measured as a function of time delay. The strongest spectral emission centered at 430 nm is independent of the time delay, while the emission centered at 550 nm is strongly dependent on the time delay, having a width of 317 fs.

2 nm). These higher-order grating resonances are a good candidate for the enhancement of our pump beam. The transmission spectra of all metasurfaces (Table S1), around both pump and signal wavelengths, is shown in Figure S2. We also measure the metasurface transmittance when it is illuminated at oblique incidence. Figure 3(c) shows the transmittance at incident angles, $\theta$, from zero to five degrees. We observe a clear splitting of the resonance, demonstrating the strong nonlocal characteristic of the metasurface.

## 3.2 | Nonlinear Characterization

Next, we measure the nonlinear emission of the metasurface, Figure 4(a), when the excitation beam is spectrally tuned to the position of the resonances. First, we start by measuring the enhancement of the SHG process when the beam is tuned to the resonances observed in the linear transmittance spectrum, Figure 3. We use a Ti:Sapphire femtosecond laser coupled to an Optical Parametric Oscillator, see Sections S2-4 of the Supporting Information. The system can be tuned in the range of 680–1600 nm, having a repetition rate of 80 MHz. The femtosecond pulses were characterized using a frequency-resolved optical gating (FROG) technique, obtaining pulse durations of $\approx 160$ fs for the $\sim 850$ nm range and $\approx 200$ fs for the $\sim 1550$ nm range (see Sec. S2 of the Supporting Information). The average laser power was kept to 10 mW. It is worth noting that the resonance width of the metasurface ($\sim 38$ nm) is similar to the bandwidth of the laser pulses, which are 23 nm in the 1550 nm range and about 9 nm in the 850 nm range. In this way, we can couple the entire energy of the laser pulses to the resonator, as the cavity rise time is faster than the pulse duration. For higher Q-factor resonances, we can experience spectral filtering and pulse broadening (see Sec. S4 of the Supporting Information).

The nonlinear enhancement of the SHG from the incident beam in the spectral range of $\sim 850$ nm is shown in Figure 4(b). In this range, the metasurface exhibits multiple higher-order resonances, as also seen in the linear transmittance spectrum, Figure 3(b). In the experiment, we scan the wavelength of the excitation laser and measure the emitted SHG from the metasurface. With this SHG spectroscopy, we obtain the wavelengths for maximum far-field enhancement due to the metasurface resonances. In Figure 4(b), we see two strong enhancement peaks at the fundamental wave (FW) of 827 nm and at 856 nm. The first resonance is not prominent in the linear spectrum, likely because it could not be spectrally resolved by our spectrometer. The second resonance, however, matches well the observed spectral dip at 860 nm. At this resonance, we measure 66 times enhancement of the nonlinear emission, in comparison to the thin film LiNbO₃. In the 1550 nm spectral range, we also scan the excitation wavelength and observe a strong emission enhancement at 765 nm, corresponding to FW of 1530 nm, Figure 4(c). This nonlinear SHG enhancement is consistent with the excitation of the metasurface bright mode at this wavelength, as also seen in the linear characterization in Figure 3(b). We note that the gaps in the scanning range are due to issues with the laser tunability. Notably, the SHG enhancement at the resonant wavelength shows a 132 times SHG enhancement as compared to the bare LiNbO₃ thin film. Overall, in both spectral ranges, the SHG dependence on the FW power has a slope close to two on the log-log scale (quadratic in power), as shown in Figure 4(d). In contrast, the measured dependence of the SFG process is linear with signal power (at 1530 nm) while keeping the pump power (at 860 nm) constant. This linearity with signal power is important for the proper transfer of the quality of gray-scale images. We also quantify the polarization dependencies of the SFG process and both SHG processes, which are shown in Figure S6 of the Supporting Information. As expected, both processes show a figure-of-eight shape when the input polarization is varied.

Next, we quantify the metasurface nonlinear emission when it is simultaneously excited by two incident beams: a pump beam at 860 nm and a signal beam at 1530 nm. These wavelengths correspond to the maximum enhancement for SHG. The two beams are derived from a femtosecond OPO. The pulses of the two beams are synchronized in time by a delay line, spatially combined by a dichroic mirror, and then focused on the metasurface by a plano-convex lens of 5 cm focal length, see Figure S3 of the Supporting Information. A $20\times$ microscope objective collected the detected SFG emission. Figure 4(e) shows the measured spectra when each of the beams excites the metasurface independently and when they overlap in time and space on the metasurface. Notably, the visible SFG emission (550 nm peak) is resonantly enhanced by the metasurface and exhibited a record enhancement of 458 times, as compared to the bare LiNbO₃ thin film. The polarization dependence for the SFG process is also tested and is shown in Figure S6 of the Supporting Information with the blue and green curves. In the first case, we fix the polarization of the signal beam along the groves of the grating and vary the polarization of the pump beam (blue curve). In the second case, we fix the polarization of the pump beam along the groves of the grating and vary the polarization of the signal beam (green curve). In both cases, we observe a figure-of-eight polarization dependence, which shows that the SFG is enhanced when both beams are co-polarized, having their polarization along the groves of the nanograting, thereby exciting the strong optical resonances and aligning with the strongest nonlinear coefficient of LiNbO₃.

Finally, in Figure 4(f), we demonstrate the time resolution of the obtained SFG. In this experiment, we varied the optical delay line between the pump and the probe pulse and measured the spectral emission from the metasurface. The SHG of the pump beam is seen in Figure 4(f) as a continuous line at 430 nm, while the SFG process is a short peak showing maximum intensity at a specific delay for which the two pulses overlap. This peak corresponds to the convolution of the two pulses and is measured to be 317 fs. This important result shows the ultra-fast speed of the SFG process, which is important for fast imaging applications, including microscopy and ultrafast vision.

In addition to the presented metasurface, we have also tested several other LiNbO$_3$ metasurfaces with higher Q-factors, on the order of 360 (see Table S1 and S2 of the Supporting Information). These nearly ten times higher Q-factor metasurfaces could result in an order of magnitude higher SFG efficiency. However, as shown in Section S4.2 of the Supporting Information, the measured SFG efficiency is only slightly enhanced. For a metasurface with a Q-factor of 360, the efficiency enhancement is only 1.3 times higher than for the metasurface with a Q-factor of 40. For a metasurface with a Q-factor of 363, the efficiency is two times higher than the metasurface with a Q-factor of 40. These measurements show that for the higher Q-factor metasurfaces, we are experiencing a frequency filtering of the signal pulse, meaning that only part of the input energy is coupled to the metasurface resonance. We also note that overall efficiency is not only proportional to the Q-factor but also to the spatial overlap of the pump and signal modes, which is why the metasurfaces with Q-factors of 360 and 363 exhibit different conversion efficiencies despite the close value of their Q-factors. Overall, when we normalize the conversion efficiency to the intensity of the pump beam, we measure an efficiency of $1.93 \times 10^{-5}$ cm$^2$/GW for the metasurface with $Q \approx 40$, which is a record value for SFG in a LiNbO$_3$ metasurface.

## 3.3 | Image up-conversion in nonlocal metasurfaces

After demonstrating the strong enhancement of the SFG process by the LiNbO$_3$ metasurface, we turn to testing its up-conversion imaging performance. As mentioned before, all spatial frequencies of the image must be equally up-converted to avoid loss of spatial resolution. This condition, however, is impeded by the nonlocality of the metasurface, as the spectral position of the resonance quickly changes with the incident angle. To circumvent the spatial dispersion of the metasurface, we take advantage of the coherent nature of the SFG upconverted process and place the metasurface up-converter in the Fourier plane of the target, rather than the image plane, as previously used [7]. Subsequently, the up-converted Fourier image of a test target is converted to the image plane of an uncooled CMOS camera (Thorlabs, DCC1545M). This is obtained by an array of plano-convex lenses, Figure 5(a), see also Figure S6 of the Supporting Information. The size of the image on the camera can be controlled by adjusting the position of the target with respect to the back-focal plane of the first lens; see detailed discussion in Section S5 of the Supporting Information. In this procedure, all spatial frequencies are homogeneously up-converted, thereby preserving the resolution of the original IR image. It is important to note that our metasurface with a periodicity of $a = 879$ nm is no longer subwavelength for the 550 nm SFG emission and displays diffraction orders. Three diffraction orders are emitted in the air, each having a different intensity. For imaging, only the zeroth diffraction order is collected and imaged into the camera. The other diffraction orders are filtered out when a $20\times$ microscope objective is used in the imaging system but they can be observed with a $50\times$ microscope objective, which has a higher numerical aperture. The intensity of the different diffraction orders can be controlled depending on the excitation direction of the metasurface. As shown in Figures S4-5 of the Supporting Information, when the metasurface is excited from the air side, the $\pm 1$ diffraction orders have the dominant intensity. However, when the metasurface is excited from the substrate side, the highest intensity is observed in the zeroth diffraction order. Therefore, we use the latter geometry in all of our measurements.

In the image up-conversion experiment, we present the images that are transformed from infrared to visible green light using the LiNbO$_3$ metasurface, as depicted in Figure 5(b). The top row of the figure displays the optical microscopy images of the target taken on a commercial optical microscope (Olympus BX51). We can see that all the details, including horizontal and vertical lines as well as numbers, are accurately reproduced in the up-converted green images. Another example of up-conversion imaging is shown in Figure 5(c), corresponding to a Siemens star target. In this target, the radial lines become closer to each other as they approach the center of the star. As such, the spatial frequencies of the target gradually increase towards the center. The Siemens star is, therefore, a convenient way to test the imaging resolution of the upconversion imaging system, which includes the joint contribution of the optical system and the intrinsic resolution of the up-converting metasurface. From the left image in Figure 5(c), we can see that the central part of the Siemens star is not resolved, obtaining spatial resolution of the system on the order of 50 $\mu$m. However, we believe that this resolution is limited mainly by the optical system rather than by the metasurface. The imaging performance of our up-conversion system is further clarified by comparison of the imaging quality with direct IR imaging by InGaAs camera and an up-converter by a thin film. The results are presented in Section S5.3 of the Supporting Information. Figures S11-S12 show the metrics for image resolution and signal-to-noise ratio (SNR) of the different IR imaging techniques. The results show that the nonlocal metasurface up-converter performs equally well as direct infrared imaging by InGaAs camera and outperforms a thin-film upconverter in resolution and SNR.

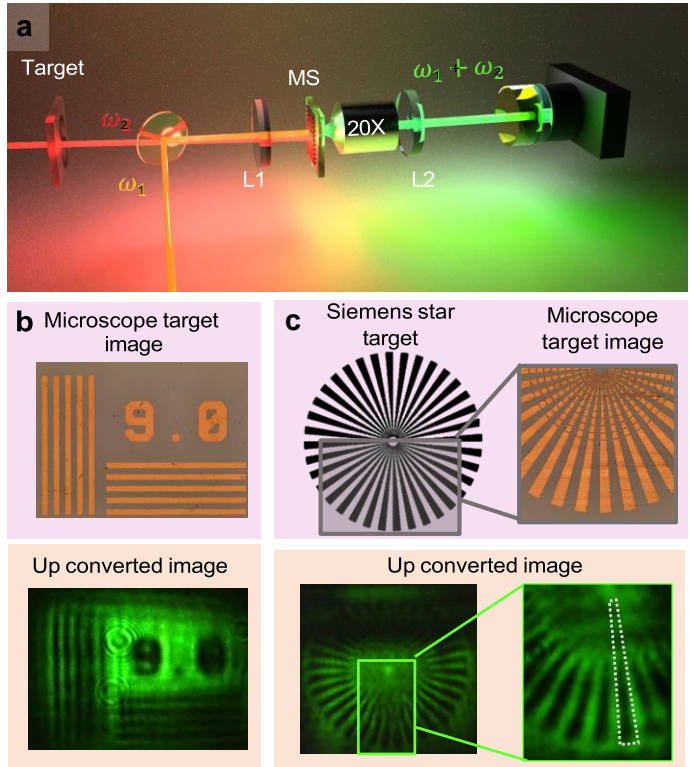

**FIGURE 5** Up-conversion imaging performed by the nonlocal metasurface. a) Schematic of optical setup used for the imaging. The signal beam illuminates a resolution target, which is Fourier imaged on the metasurface. Mixing the IR image with the pump beam results in a visible image of the target (in the SFG), which is subsequently imaged onto a conventional CMOS visible camera. (b,c) Up-converted images of different parts of the target. Optical microscope images of the resolution target (top) and their corresponding up-converted images (bottom). In (b,) the horizontal and vertical stripes are 50 $\mu$m wide. In (c), the imaging of the Siemens star allows for estimation of the resolution of the optical system, which is up to 50 $\mu$m in the upconverted images.

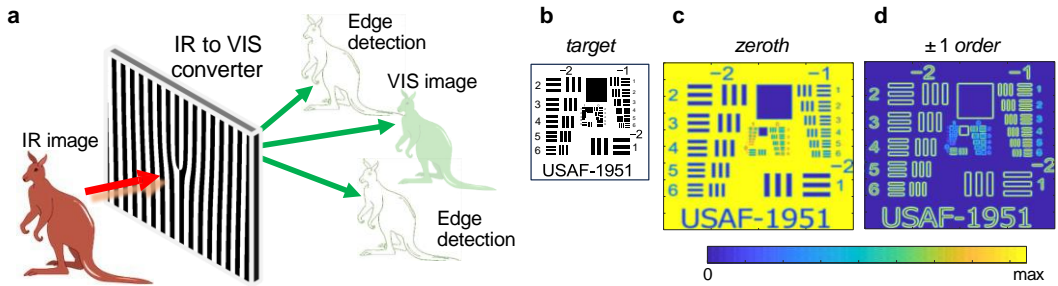

**FIGURE 6** Concept of combined imaging and image processing using spatially-variant nonlocal metasurfaces. a) Schematic of the image processing, where a nonlocal metasurface with phase dislocation can generate a direct image in the zeroth order, together with edge-detection of the image in the $\pm$ first diffraction order. (b-d) Numerical demonstration based on USAF-1951 imaging target. b) Image of the resolution target (ground truth). c) Direct up-converted image in the zeroth order, following Eq. (2). d) Calculated up-conversion imaging in the $\pm 1$ orders using the convolution theorem; see the Supporting Information.

Overall, we demonstrate that it is possible to convert IR images into visible in an ultra-thin metasurface through the resonantly enhanced nonlinear process of SFG. We obtain high-resolution images of the target despite the dispersive nature of the nonlocal metasurface.

## 4 | DISCUSSION AND CONCLUSIONS

The demonstrated up-conversion imaging with high spatial resolution is underpinned by the coherent nature of the SFG up-conversion process with Fourier space up-conversion geometry, thereby overcoming the nonlocal character of the nonlinear enhancement. Its high efficiency is further enabled by the high-Q factor resonances in the nonlocal metasurface and the high transparency of the material. The measured efficiency is an order of magnitude higher than previous results [7] and exceeds many results obtained with large bulk crystals [3]. Further improvements could be envisaged by the use of transparent materials with high nonlinear coefficients, such as GaP [10] or InGaP [45]. The exploration of inverse design techniques for band engineering and polarization control [46] could allow for further quality improvement and polarization independence response based on 2D metasurface arrays.

Importantly, the metasurface up-conversion further offers unique opportunities for simultaneous up-conversion and image processing that is not accessible in bulk crystals. For example, different imaging functions can be encoded in the different diffraction orders of the nonlinear emission, see Figures S4-5 of the Supporting Information. For example, the different nonlinear-diffraction orders can carry information about the polarization state of the light [14] or can perform different image processing operations [16], in addition to the up-conversion process. Such image processing could, for example, be enabled by spatially variant nonlocal metasurface [47, 48], which can perform direct up-conversion imaging in the zeroth diffraction order while performing edge-detection in the first diffraction orders, Figure 6(a). In this example, we use a 1D nonlocal metagrating with a topological dislocation. Such dislocation is expected to have little effect on the direct SFG emission; however, it can induce a substantial edge-detection effect in the $\pm 1$ diffraction orders of the SFG emission. These are derived from the convolution of the signal beam (image of the target), the Gaussian pump beam, and the grating function, see Sec. S5.3 of the Supporting Information. Figure 6(b) shows the ground truth image of the resolution target, and Figure 6(c) shows the calculated image in the zeroth SFG diffraction order, which resolution is affected only by the convolution with the Gaussian pump beam. In contrast, the $\pm 1$ SFG orders are derived from the convolution of the signal target image, the Gaussian pump beam, and the grating function, thereby producing edge detection operation of the original image, Figure 6(d). In this case, the smallest features are not resolved due to the finite size of the pump beam.

In conclusion, we have experimentally demonstrated enhanced IR to visible up-conversion imaging using a high-Q resonant nonlocal metasurface. We have achieved a record high up-conversion imaging efficiency with high spatial resolution. The high spatial resolution is achieved despite the strong nonlocality of the metasurface by exploiting the coherent up-conversion nature of the SFG process through Fourier-plane upconversion. We have further alluded to the unique image-processing functionalities offered by the up-converting metasurface to perform direct imaging and edge detection simultaneously in a single-shot setting. We believe that our study could find important applications in future compact night vision instruments, sensor devices, and multi-color imaging at room temperature.

## Acknowledgements

We acknowledge useful discussions with Costantino De Angelis, Jinyong Ma, and Neuton Li.

## Conflict of Interest

The authors declare no conflict of interest.

## Supporting Information

Supporting Information is available.

## GRAPHICAL ABSTRACT

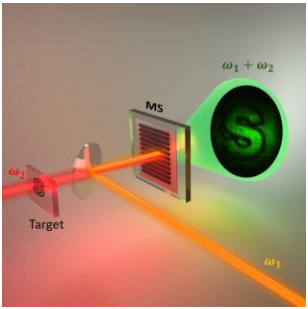